\documentclass[conference]{IEEEtran}
\IEEEoverridecommandlockouts

\usepackage{cite}
\usepackage{amsmath,amssymb,amsfonts}
\usepackage{algorithmic}
\usepackage{graphicx}
\usepackage{textcomp}
\usepackage{xcolor}
\def\BibTeX{{\rm B\kern-.05em{\sc i\kern-.025em b}\kern-.08em
    T\kern-.1667em\lower.7ex\hbox{E}\kern-.125emX}}

\usepackage{url,hyperref}

    \makeatletter
    \newcommand{\linebreakand}{%
      \end{@IEEEauthorhalign}
      \hfill\mbox{}\par
      \mbox{}\hfill\begin{@IEEEauthorhalign}
    }
    \makeatother

\begin{document} 

\title{(No) Separation of Concerns: Philosophical Concerns about Agent Ethics and Implications for SE Practitioners}

\author{\IEEEauthorblockN{Marc Cheong\IEEEauthorrefmark{1}}
\IEEEauthorblockA{\textit{The University of Melbourne}\\
Parkville, Australia \\
marc.cheong@unimelb.edu.au}
\and
\IEEEauthorblockN{Sebastian Baltes}
\IEEEauthorblockA{\textit{Ruprecht-Karls-Universität Heidelberg}\\
Heidelberg, Germany \\
sebastian.baltes@uni-heidelberg.de}
\and
\IEEEauthorblockN{Joanne Byrne}
\IEEEauthorblockA{\textit{The University of Melbourne}\\
Parkville, Australia \\
joanne.byrne@unimelb.edu.au}
\linebreakand
\IEEEauthorblockN{Haoyu Gao}
\IEEEauthorblockA{\textit{The University of Melbourne}\\
Parkville, Australia \\
haoyug1@student.unimelb.edu.au}
\and
\IEEEauthorblockN{Raula Gaikovina Kula}
\IEEEauthorblockA{\textit{University of Osaka} \\
Osaka, Japan \\
raula-k@ist.osaka-u.ac.jp}
\and
\IEEEauthorblockN{Sarita Rosenstock}
\IEEEauthorblockA{\textit{The University of Melbourne}\\
Parkville, Australia \\
sarita.rosenstock@unimelb.edu.au}
\linebreakand
\IEEEauthorblockN{Christoph Treude}
\IEEEauthorblockA{\textit{Singapore Management University} \\
Singapore \\
ctreude@smu.edu.sg}
\and
\IEEEauthorblockN{Mansooreh Zahedi}
\IEEEauthorblockA{\textit{The University of Melbourne}\\
Parkville, Australia \\
mansooreh.zahedi@unimelb.edu.au}
}

\maketitle
    \begingroup
    \renewcommand\thefootnote{}
    \footnotetext{\IEEEauthorrefmark{1}MC is corresponding author. Other authors are listed in alphabetical order.}
    \endgroup
\thispagestyle{plain}
\pagestyle{plain}

\begin{abstract}
The use of AI agents in software engineering (SE) are commonplace, with differing levels of autonomy, a myriad of capabilities, and ability to perform a variety of tasks. Philosophers have begun to examine the ethics of such agents, and SE researchers or practitioners could benefit from them; on the other hand, SE practitioners currently explore most of these concerns through the lens of principlism. We bridge this gap by reviewing the philosophical literature on AI agents indexed in PhilPapers, and translating  its ethical concerns into a research agenda for SE researchers and into implications for teams that use coding agents. A systematic mapping study of PhilPapers -- a major philosophy index and open access archive -- yielded 33 peer-reviewed papers by professional philosophers published since 2023. We coded these papers thematically, with cross-disciplinary review of the emerging themes, resulting in eight themes and 85 codes. From these, we derive seven research questions for SE, covering developers' reliance on agents, responsibility for agent-written code, ethical constraints in agent configurations, and benchmark design. For practitioners, we derive five implications: decide which choices an agent must refer to a human, keep agent-written pull requests small enough to review, pair configuration-file instructions with checks, watch for changes to tests, and record who approved what. We argue that ethical concerns cannot be separated from SE concerns into a module of their own: they arise inside code review, configuration, evaluation, and requirements work, and addressing them requires cross-disciplinary work.
\end{abstract}


\textbf{\color{red} This work has been submitted to the IEEE for possible publication. Copyright may be transferred without notice, after which this version may no longer be accessible.
}
\section{Introduction}

In software engineering (SE), AI agents are moving from merely suggesting code to acting on their own in SE workflows: from writing/merging/deploying code, to maintaining documentation, to triaging issues and auditing repos  \cite{Treude2026, Liu2026}. Concerns about the socioethical impact of AI agents (\textit{viz.} the  \textit{agentic AI} paradigm) are beginning to receive attention from the SE community \cite{aleti2026trustworthyaisoftwareengineers,gao2026ethicsconfigurationaiethics}, especially in the \textsc{Agent} workshop series \cite{HODA2026,Assalaarachchi2026}.

Many of the choices these agents make while completing a task are ethical choices, although they look like implementation details. An agent that adds a feature also decides how long the feature keeps user data, whether personal data is written to logs, and whether a new screen works with a screen reader. When the agent opens the pull request and a developer approves it, nobody may have made these choices deliberately.

Social scientists and philosophers have begun to unpack concerns about AI agents at length, against the broader backdrop of \textit{digital ethics} and \textit{AI ethics}\footnote{See e.g., \url{https://philpapers.org/browse/ethics-of-artificial-intelligence}.}. \textit{Translational} work -- which brings together philosophers, social scientists, and SE practitioners -- in order to address philosophical concerns about AI agents into SE terms, is few and far between. 

It is in this spirit of bridging the disciplinary gap that we authored this paper: with the aim of reviewing the extant philosophical literature on AI agents and translating its ethical concerns into a research agenda for software engineering. We posit two core research questions in our work:
    \begin{itemize}
        \item \textbf{RQ1}: What ethical concerns about AI agents does the philosophical literature raise?
        \item \textbf{RQ2}: How do these concerns map onto SE research and practice, given the increasing adoption and pervasiveness of agents?
    \end{itemize}
    
To answer these \textbf{RQs}, our contributions take the form of a thematic map of ethical concerns drawn from a systematic survey of PhilPapers, ``a comprehensive index and bibliography of philosophy maintained by the community of philosophers''\cite{ThePhilPapersFoundation2026} (\textbf{RQ1}), together with a research agenda of seven questions and five implications for SE practice (\textbf{RQ2}), both developed by a diverse author team, from all career stages. This paper is structured as follows: Section \ref{s:Background} is a preliminary overview of related literature; Section \ref{s:Method} describes our method; Section \ref{s:Results} presents our results in response to \textbf{RQ1} and \textbf{RQ2}; Section \ref{s:Discussion} discusses the results, states implications for SE practice, and lists threats to validity; and Section \ref{s:Conclusion} concludes.

%%%%%%%%%%%%%%%%%%%%%%%%%%%%%%%%%%%%%%%%%%%%%%%%%%%
%%%%%%%%%%%%%%%%%%%%%%%%%%%%%%%%%%%%%%%%%%%%%%%%%%%

\section{Background and Related Work}\label{s:Background}
\subsection{Agentic AI vs the concept of Agency}\label{ss:Agentic}
To borrow from the \textsc{Agent 2027} workshop, Agentic AI systems in SE ``exhibit goal-directed autonomy, reasoning, tool use, interaction, and continuous evolution''\footnote{\url{https://conf.researchr.org/home/icse-2027/agent-2027}}. This aligns with the popular conception of AI agents (``Autonomous software programs that perform specific tasks'' \cite{sapkota2026}) and agentic AI as-a-whole (``Systems of multiple AI agents collaborating to achieve complex goals'' \cite{sapkota2026}). However, it is important for us to raise the fact that the term ``deserves further specification'' \cite{Ranisch2026}: \textit{agency} means something rather different in philosophy and ethics. Put simply, in this second context, \textit{agency} ``...denotes the exercise or manifestation of this capacity [to act]'' \cite{sep-agency}, and philosophers are concerned with issues such as ``action in terms of intentionality... [and also] causation by the agent's mental states and events'' \cite{sep-agency}.

Hence, this paper's review of the philosophical literature seeks to capture both concerns. The first, as Mussgnug aptly positions it \cite{Mussgnug2025} -- ``as a technical term encountered throughout current public discourse, rather [than] signaling any philosophical notion of moral or epistemic agency'' -- as well as the second, which incorporates the moral/epistemic perspective. This turns out to be an integral part of setting out our research agenda (Section \ref{ss:ResearchAgenda}) and our implications for SE practitioners (Section \ref{ss:Implications}).

\subsection{Ethics, governance, and responsibility in agentic SE}
Human factors, including ethics/governance/societal impact of AI use, has been encapsulated as part of frameworks, principles, and guidelines  \cite{Hagendorff2020,Jobin2019}. In terms of collaboration in SE, Treude and Gerosa \cite{Treude2025} taxonomised the interaction between human and AI.

Specifically for agentic SE, Hoda \cite{HODA2026} proposed CRAFT (comprehensive, responsible, adaptive, foundational, translational) values for agentic SE ``that extends the focus of agent design beyond coding to human–AI collaboration'' (as cited in \cite{Treude2026}). Aleti and colleagues worked on addressing ``trustworthiness of AI software engineers'', from a myriad of dimensions including  ``technical quality, transparency and accountability, epistemic humility, and societal and ethical alignment'' \cite{aleti2026trustworthyaisoftwareengineers}. Prior research by Treude has also examined ``ownership, responsibility, liability, and disclosure obligations'' between AI coding tool providers and software developers \cite{TreudeTOS2026}.

When it comes to agents in SE, repository-level context files (in particular \texttt{AGENTS.md}) have been identified as a potential bearer of ethical instruction and constraint \cite{Treude2026,gao2026ethicsconfigurationaiethics}. It is important to understand how ethical considerations are codified in these artefacts, which should (in theory) guide agent behaviour in the SE process. As a case in point, Gao and colleagues \cite{gao2026ethicsconfigurationaiethics} looked into AGENTS.md and related artefacts to find prevalences of ethical statements aligning with Australia's AI Ethics Principles.

\subsection{Going beyond current principles-based approaches}

What sets our work apart from extant ones is that we \textit{go beyond principlism}. To wit: current approaches mainly draw upon established principles in applied AI ethics. To paraphrase Cheong and Coghlan \cite{Cheong2025-uj}: principlism has ``achieved a high profile in digital ethics [where it is]... an attempt to distil insights from moral theory and philosophy[...] to produce a guiding framework that different [...] theorists can at least to some extent agree on''. Principlism in AI ethics can trace its inspiration to applied medical ethics \cite{Beauchamp2001-qs}, where it has helped medical professionals make ethical judgement calls and choices.

While SE practitioners in general are familiar with ethical principles \cite{Pant2024,Gao2024}, they faced ``general, technology-related, and human-related challenges in applying these principles'' (as cited by \cite{gao2026ethicsconfigurationaiethics}). This is not uncommon in the study of applied ethics: principles might conflict with one another, leaving one in a conundrum as to which principle takes precedence \cite{Cheong2025-uj}. Also, focusing on a plurality of principles \cite{Jobin2019} will neglect  other ethico-philosophical traditions, including utilitarian ethics and duty ethics, to name but a few examples. 
Therefore, this paper intends to address the translation gap (across philosophy and SE) as well as the framework/theory gap (going beyond principlism).

\begin{figure}[!h]
    \centering
    \includegraphics[width=\columnwidth]{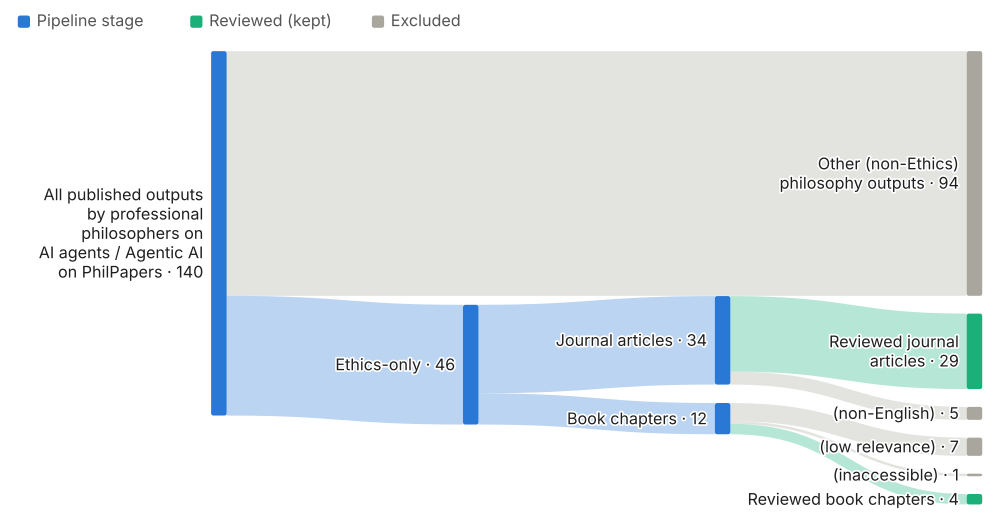}
    \caption{Sankey diagram illustrating the paper filtering and review process.  The proportion of other (non-ethics-related) philosophy of outputs -- 94 papers -- are provided solely for comparison. [Figure generated with Claude \textsc{Opus-5.5-Extra}].}
    \label{fig:Sankey}
\end{figure}

%%%%%%%%%%%%%%%%%%%%%%%%%%%%%%%%%%%%%%%%%%%%%%%%%%%
%%%%%%%%%%%%%%%%%%%%%%%%%%%%%%%%%%%%%%%%%%%%%%%%%%%

\section{Methodology}\label{s:Method}
In this section, we outline our methods: the process of searching PhilPapers and reviewing relevant papers,  qualitative coding, and refinement of themes.

\subsection{Review of PhilPapers}
PhilPapers \cite{ThePhilPapersFoundation2026} is a comprehensive online research index for academic philosophy and open-access archive of philosophy papers, i.e.,  the rough equivalent of the combination of arXiv and the DBLP computer science bibliography. It currently indexes over 3.1M+ entries in 6k+ categories, with over 150k preprints/manuscripts/open access papers hosted under its PhilArchive repository, which claims to be the ``largest open access e-print archive in philosophy\footnote{\url{https://philarchive.org/}.}.

We searched PhilPapers on 1 October 2026 for relevant papers on AI agents and the concept of Agentic AI, given the nuances in definition per Section \ref{ss:Agentic}. The following search criteria were used:

\begin{itemize}
    \item \textbf{Search string:} \texttt{("ai agent" | "agentic ai") \& (ethic*)}
    \item \textbf{Date published:} 2023 -- present; chosen in order to include publications after the public introduction of ChatGPT (as the first mass-marketed generative AI platform).
    \item \textbf{Author type:}  professional authors only. PhilPapers considers an author \textit{professional} if they either hold a doctorate in philosophy or have authored an article in one of the 71 philosophy journals on its most-popular list\footnote{\url{https://philpapers.org/help/faq.html} and \url{https://philpapers.org/journals?listId=1}}
    \item \textbf{Paper type:} published only. Similar to preprint servers (e.g., arXiv), PhilPapers contains both peer-reviewed and non-peer-reviewed preprints. The \textit{published only} filter excludes unpublished or non-peer-reviewed manuscripts and drafts, and returns peer-reviewed papers with links to the final version on the publishers page. These filters restrict the corpus to published work by credentialed philosophers, trading some coverage for quality. 
\end{itemize}

To give us context as to how many articles in PhilPapers (i.e., in \textit{all} branches of philosophy) which dealt with the concept of AI agents/agentic AI, we ran an initial search without the \texttt{ethic*} wildcard. As of time of writing, this figure amounted to $140$ papers (Figure \ref{fig:Sankey}). Focusing on \texttt{ethic*} related papers per our original intention, we obtained an initial set of $46$ papers. The papers were further filtered with the following set of inclusion/exclusion criteria. 
\begin{itemize}
    \item \textbf{Inclusion criteria:} journal articles or book chapters; peer-reviewed; currently available (as some works are forthcoming and have no online-first copy); in English; 
     \item \textbf{Exclusion criteria:} low relevance works which do not substantively addresses ethical concerns of AI agents, such as technical manuals or guides with only cursory mention of ethics.
\end{itemize}

Finally, a total of $N=33$ papers were obtained post-filtering (available in the replication package: Section \ref{s:DataAvailability}). To maximise rigour, the search process above adopted best practices for \textit{systematic mapping studies} in SE \cite{WohlinSLR2024,Kitchenham2011}. However, we do not claim completeness here, as the selection of the PhilPapers database (and the search parameters themselves) can be considered a form of convenience sampling. Differentiating our work from systematic literature reviews, the quality-assessment criteria  advocated in Kitchenham et al. \cite{KITCHENHAM2009} were relaxed in our study, as philosophical literature does not lend itself to judgment calls of "quality/validity" compared to (empirical) STEM literature. That being said, our exclusion criteria is a `gate' to omit works with low philosophical content or engagement with agentic AI ethics.

\subsection{Coding and Discussion with co-authors}
The $N=33$ papers were then coded, using an iterative qualitative coding approach. The first author performed open coding on all the papers, using the \textsc{ATLAS.ti} software package (\textsc{Web version 9.26.2-2026-09-29}), to identify relevant passages and initial codes. The coding process focuses on passages concerning ethics, philosophy, morality, responsibility, or governance of agentic AI or AI agents. Generic discussion on GenAI or LLMs outside the context of agents are excluded; however, discussion of LLMs as bots or agents (e.g., medical chatbot, or customer service agent) is included as it fits our definition outlined in Section \ref{ss:Agentic}.

Following  iterative coding (Section \ref{ss:Thematic}), the list of initial themes and codes were reviewed by three authors (JB, SR, SB), who collectively have expertise in philosophy (especially digital ethics), anthropology, and human aspects in software engineering. Feedback from this process has informed the final hierarchy of themes and sub-themes (codes). To assist coauthors with inspecting, visualising, and parsing the themes/codes, Claude \textsc{Opus-5.5-Extra} was used to generate a \textit{kanban}-like interface (available in the replication package: Section \ref{s:DataAvailability}).

%%%%%%%%%%%%%%%%%%%%%%%%%%%%%%%%%%%%%%%%%%%%%%%%%%%
%%%%%%%%%%%%%%%%%%%%%%%%%%%%%%%%%%%%%%%%%%%%%%%%%%%

\section{Results}\label{s:Results} 
\begin{figure*}[!htb]
    \centering
    \includegraphics[width=.8\linewidth]{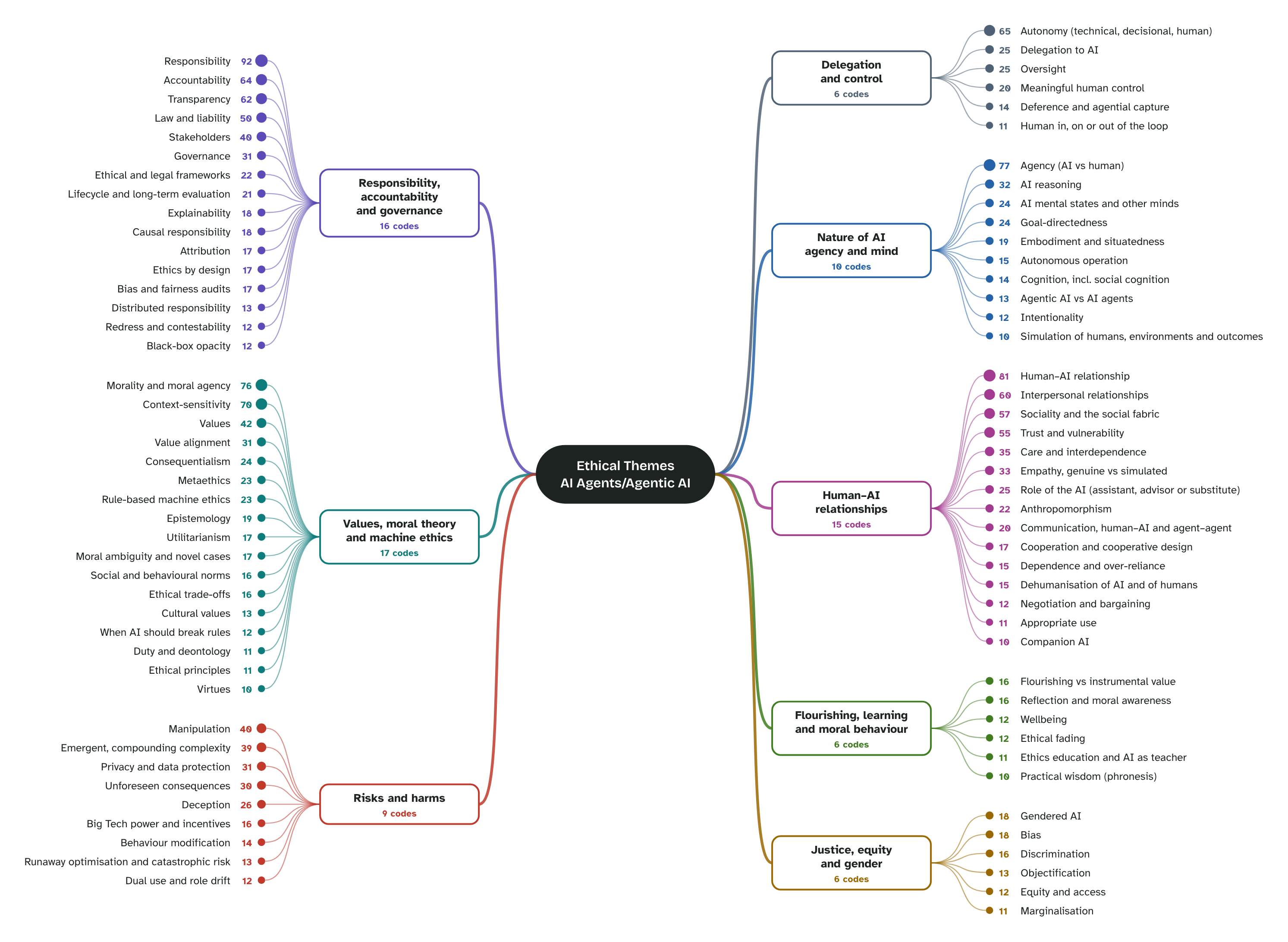}
    \caption{Thematic map listing of key themes and codes found from our review (total papers = 33). [Figure generated with Claude \textsc{Opus-5.5-Extra}].}
    \label{fig:thematicmap}
\end{figure*}

\subsection{Thematic map}\label{ss:Thematic}

Initially, a total of $208$ initial codes were generated from $847$ passages (non-mutually-exclusive), including in-vivo direct quotes which contextualise key definitions in the discourse \cite{InVivoCoding2008}. These initial codes are then arranged based on thematic clusters, with minor codes merged or dropped. After two iterations, we arrive at a list of $8$ high-level themes, with a second level of $85$ codes (or sub-themes).

Figure \ref{fig:thematicmap} illustrates the key themes and codes found in our review. Each major theme (e.g., \textit{Nature of AI agency and mind}) consists of second-level codes (e.g., \textit{Embodiment and situatedness}); the number next to each code denotes the number of passages linked to it. Note that the number of passages is not indicative of code prevalence, as different papers have different passage lengths. It is important to note that this taxonomy of themes (codes) arose from the $33$ papers, and by no means a `be-all-and-end-all' hierarchy of ethical concerns regarding AI agents. Prior studies of ethical principles in applied AI ethics reveal a plurality of permutations \cite{Jobin2019}, as well as boundary cases straddling the definition of several principles \cite{gao2026ethicsconfigurationaiethics}. In no particular order, an overview of these $8$ top-level themes follows, along with characteristic quotes from the literature describing  concerns by philosophers (with \textit{emphasis added)}.

\subsubsection{Responsibility, accountability and governance} 
``\textit{Legal systems are ill-prepared to address agentic AI}. Questions of accountability are especially tricky when addressing the \textit{layered and complex developer ecosystem} underpinning today's agentic AI systems...'' \cite{Tweed2027};     ``The main issue here is the level of control over the outcome. In fully controllable systems, responsibility is clearer, while \textit{in systems with a high degree of unpredictability, such as autonomous agents, determining responsibility becomes challenging}...'' \cite{Lavazza2026}.

\subsubsection{Values, moral theory and machine ethics} 
    ``Realizing the goal of value alignment requires taking a substantive position on \textit{what value judgments should ultimately anchor the behavior of AI agents}...'' \cite{Riesen2025}; 
    ``In addition, \textit{strict norms }against deception and manipulation that are unenforceable for humans \textit{might be enforceable in AI}.'' \cite{Tarsney2025}  
    
\subsubsection{Risks and harms} 
    ``AI Agents and Agentic AI ... require almost unrestricted access to sensitive personal data... which already touches on the ethical implications of these systems, such as those related to \textit{privacy and transparency}. Their capacity to act with a high degree of technical autonomy and with minimal to no human intervention underscores both their enormous potential and their \textit{ambivalence and raises questions regarding responsibility, fairness and non-maleficence}.'' \cite{Hahn2026}; 
    ``Human beings will need to \textit{carefully oversee and check} the work of AI agents to catch and correct errors. In some cases, it may take \textit{more effort to find and fix the problems} created by an AI agent than to use existing alternatives...'' \cite{Hosseini2026}.

\subsubsection{Delegation and control} 
    ``At this point, it is necessary to distinguish between the technical autonomy and the decisional autonomy of AI systems, as a system may be \textit{technically autonomous while remaining merely advisory} in decisional terms...''  \cite{Gounaris2026}; 
    ``As AI Agents and Agentic AI systems possess a \textit{higher degree of agency, systemic interconnection and technical autonomy}, the issue of whether a \textit{human remains in the loop}, on the loop, or out of the loop becomes even more pressing compared to traditional AI systems.'' \cite{Hahn2026}

\subsubsection{Nature of AI agency and mind} 
    `The questions of moral, ethical, and legal responsibilities in AI presuppose that AIs are intelligent and can execute actions that could be \textit{blame-worthy (or praise-worthy)}. If thus, then they are intelligent systems, otherwise known as agents. \textit{If they are agents, it presupposes they have or could have autonomy}. Hence, the need to expound the notion of "action" and "agency" in AI...' \cite{Onyeukaziri2023}
    ``The results show that users selectively transfer responsibility to AI, with a strong preference for\textit{ cognitively oriented tasks} such as researching information, translation, and study/work-related support...'' \cite{Hasselwander2026}
    
\subsubsection{Human–AI relationships}  
    `This iterative, trust-based engagement is fundamentally undermined when some “participants” may be \textit{AI systems incapable of genuine transformation or human vulnerability}...' \cite{Burns2025};
    ``Trust is not intended to be about whether an agent should trust a given extended resource but whether they do in fact exhibit trust in it. This condition was originally known as \textit{automatic endorsement, i.e., systems are trusted if their deliverances are treated with tacit concordance} and without scrutiny.'' \cite{Clowes2026}
    
\subsubsection{Flourishing, learning and moral behaviour}  
    ``cultivation of phronesis (practical wisdom), a capacity to be able to render the optimal decisions across situations leading to one’s flourishing... although AI might be able to provide more accurate advice for reasoning and decision-making ... their lack of emotional and  behavioral capacities may significantly \textit{limit their ability to interact with and stimulate students for their phronesis cultivation}'' \cite{Han2026};
    ``Given that AI chatbots represent an inherently ambiguous identity, people may similarly \textit{disengage from the ethical standards that typically govern human interactions}...  less concerned about being judged from AI chatbots, reducing impression management concerns and self-monitoring behavior that normally constrains unethical behavior'' \cite{Weirup2026}
    
\subsubsection{Justice, equity and gender} 
    ``The use of female AI agents, such as vocal assistants, chatbots and robots, is on the rise, but the indiscriminate feminization of these AI agents poses novel \textit{ethical concerns about their impact on gender relations in society}.'' \cite{Borau_2025};
    ``When AI Agents or Agentic AI systems act technically autonomously ... the ability of humans to detect potential biases is even more reduced compared to traditional AI applications. At high levels of automation, users may not even notice if an \textit{AI system systematically disadvantages certain marginalized groups} while favouring others...'' \cite{Hahn2026}

\vspace{6pt}
At first glance, responsibility and transparency (under \textit{Responsibility, accountability and governance}) are two of the five principles that AI ethics guidelines converge on, per the meta-analysis by Jobin et al. \cite{Jobin2019}. Of the remaining three convergent guidelines, non-maleficence (doing no harm) and privacy are reported under \textit{Risks and Harms}; and justice and fairness maps to our theme of \textit{Justice, equity, and gender}. Interestingly, the usage of guidelines, i.e., a form of ethical principlism, is a code in itself under \textit{Values, moral theory, and machine ethics}.

\subsection{Research agenda for Ethics in SE}\label{ss:ResearchAgenda}
Following this initial mapping, the authors translated these emergent ethico-philosophical concerns into a research agenda for SE. Each question names the themes and codes it draws on and the SE practice in which the concern arises. Section \ref{ss:Implications} states what teams can do while the questions are open.

\subsubsection{Does sustained reliance on coding agents change developers' ability or willingness to recognise and address ethical issues?}
This spans multiple themes: \textit{Human–AI relationships} (dependence and over-reliance), \textit{Delegation and control} (deference and agential capture), and \textit{Flourishing, learning and moral behaviour} (ethical fading + practical wisdom). We suggest that SE researchers investigate whether developers become less likely to identify ethical issues when working with coding agents, and whether particular development practices or interfaces can help preserve independent ethical judgment. Furthermore, on the concept of ethical fading -- ``individuals fail[ing] to recognize the moral aspects of a situation'' \cite{Weirup2026} -- agents may make value-laden decisions without flagging them. Examples include: how long to keep data, whether to log personal data, and how accessible a UI should be. When do developers stop noticing these decisions, and when should an agent ask instead?

\subsubsection{Who maintains control and responsibility for agent-written code?} This links to the broad array of codes under \textit{Responsibility, accountability and governance}, as well as \textit{Delegation and control} (specifically: Autonomy (technical, decisional, human)). In the agentic SE context, is approving a large pull request written by an agent meaningful control in the sense of Santoni de Sio and van den Hoven \cite{SantoniDeSio2018}? One paper in the corpus refers to the SAE levels of driving automation, under which, according to that paper, responsibility is "shifted or distributed" \cite{Hahn2026} depending on the degree of automation. What would such levels look like for coding agents? Do attribution markers such as "Co-authored-by" lines change how carefully people review agent-written changes?

\subsubsection{How can ethical principles expressed in agent configurations be translated into verifiable behavioural constraints?}
Prior work \cite{gao2026ethicsconfigurationaiethics,Treude2026} has discussed the potential for ethical principles to be embedded in configuration artifacts such as \texttt{AGENTS.md}, where we already have initial evidence that developers encode ethical principles in agent configuration files, but we don't know whether agents actually follow these instructions. These principles span the gamut of codes under the broad theme of \textit{Values, moral theory, and machine ethics}: this includes, specifically, value alignment, rule-based machine ethics, ethical trade-offs, and ethics by design. We suggest future work on benchmarks for ethical instruction adherence, techniques for detecting conflicting instructions, and mechanisms for enforcing ethical constraints or escalating trade-offs to humans.

\subsubsection{When should agents follow (or break) rules in the exercising of ethical/moral judgment?} One paper in our corpus has argued that agents should break their designers' rules when the harm is low and the pro-social gain is high \cite{Ramanayake2023}. This relates to the code \textit{When AI should break rules} under \textit{Values, moral theory and machine ethics}. An interesting research question thus arises: Which ethical requirements can be written as rules in \texttt{AGENTS.md} files, and when should an agent depart from them? For example, an \texttt{AGENTS.md} file restricts an agent to one directory, and the agent finds a hard-coded password in a file outside that directory. Should it fix the file, report it, or stay within its instructions?

\subsubsection{What are we optimising for in simulations and benchmarks, and how do we even quantitatively `optimise' for ethics?}
Floridi et al's paper in our review \cite{Floridi2026} pointed out a gap between simulated benchmarking environments and real settings, in terms of ``the complexities that agents encounter in authentic digital settings''. Coding agents, too, are commonly evaluated on benchmarks such as SWE-bench. Which ethically relevant behavior, such as reporting results honestly, can a benchmark of this kind measure at all? Another deeper question concerns specification gaming, which crops up in multiple codes under \textit{Risks and harms}. Given Von Arx et al. and METR's \cite{metr-2025-recent-reward-hacking} reports that frontier models modified tests or scoring code to raise their scores, is modifying tests to raise a score deception in the philosophical sense, especially if it has other (utilitarian) benefits to balance this deceit?

\subsubsection{Does describing agent actions in intentional terms (such as "the agent decided") shift perceived responsibility away from vendors and deployers?}  A vignette experiment could test this question, which is of interest to SE practitioners looking into facets of \textit{human} (developer)\textit{--AI} (agent) \textit{relationships} (anthropomorphism);  \textit{Nature of AI agency and mind} (AI mental states); and \textit{Flourishing, learning, and moral behaviour} (ethical fading \cite{Weirup2026}).

\subsubsection{Whose values and interests are represented when coding agents make requirements and design decisions?}
This would be mostly related to codes within \textit{Justice, equity and gender} (marginalisation/discrimination), \textit{Values, moral theory and machine ethics} (cultural values/social and behavioural norms), and \textit{Responsibility, accountability, and governance} (stakeholders). We suggest to investigate whether agents systematically overlook certain stakeholders, whether AI-generated personas represent real users well enough, and whether reliance on simulated stakeholders reduces participation by actual stakeholders. This could lead to empirical studies comparing human and AI-supported requirements engineering approaches, for example.

%%%%%%%%%%%%%%%%%%

\section{Discussion}\label{s:Discussion} 
Reflecting upon our results, the ethical concerns that philosophers raise about AI agents do not sit within a clear delineation; consider that four of {the seven questions} on our research agenda draw on two or more themes at once, and the agenda as a whole touches everyday SE activities such as code review, configuration, evaluation, and requirements engineering.

In answer to \textbf{RQ1}, our review of $33$ papers yielded eight themes and $85$ codes (Figure~\ref{fig:thematicmap}), ranging from responsibility and governance to justice, equity, and gender; many of which align with extant principles-based approaches in AI/digital ethics \cite{Jobin2019}, and the convergent principles AI ethics guidelines mentioned at the end of Section \ref{ss:Thematic}. In our answer to \textbf{RQ2}, we translated these themes into seven research questions (Section~\ref{ss:ResearchAgenda}), anchored where possible in concrete SE practices: reviewing and approving agent-written pull requests, writing \texttt{AGENTS.md} files, benchmarking coding agents,  eliciting requirements, and other core SE tasks.

As alluded to in our Background (Section \ref{ss:Agentic}), the two perspectives diverge most clearly on the concept of \textit{agency} itself. SE uses ``agent'' as a technical term for software that acts on its own~\cite{Mussgnug2025}, whereas philosophers tie agency to intentionality and mental states, as can be seen from our thematic map in Figure \ref{fig:thematicmap}. 

Some concepts, as is expected, did not translate cleanly, or have the same relevance/weight, as its nearest SE terms. For instance, practical wisdom (\textit{phronesis}) and flourishing~\cite{Han2026} largely concerns agents in social and educational roles with end users, whereas coding agents mostly act on SE artifacts. These and similar concepts also resist being expressed as requirements, rules or benchmark metrics, which is how SE usually operationalises its concerns (cf.\ Section~\ref{ss:ResearchAgenda}). 

From the authors' experience in SE, the codes that are most directly applicable are not the ones with the most passages: ethical fading~\cite{Weirup2026}, deference and agential capture (in essence: the overriding, deferring, and subsuming of ethical evaluation and responsibility), dual use and role drift, and when AI should break rules. Ethical fading and deference and agential capture underlie the first two implications below.

\subsection{Implications for SE practice}\label{ss:Implications}
The questions in Section~\ref{ss:ResearchAgenda} are open, but teams that use coding agents already make the decisions behind them. We state five implications of the reviewed literature for such teams. They correspond to the first, second, third, fifth, and sixth questions on our agenda. They follow from philosophical arguments about AI agents in general, and none has been tested with SE teams.

\subsubsection{Choices referred to a human} Ethical fading, originally discussed by Tenbrunsel and Messick \cite{tenbrunsel2004ethical}, describes how individuals can overlook the moral aspects of their decisions. This phenomenon has also been studied in human–AI interactions \cite{Weirup2026}. In coding-agent workflows, it could occur when an agent picks a default and the diff shows only code. Teams can name the types of choices that an agent must not make alone, and instruct the agent to stop and ask when a task involves one. In addition, the pull request description should list every choice that the task did not specify.

\subsubsection{Review of agent-written pull requests} Approving a pull request is not the same as controlling it. Clowes \cite{Clowes2026} examines how AI systems can become incorporated into human cognitive processes. A related concern in SE is that reviewers may accept agent-generated changes without sufficiently scrutinising them.  A team that counts approval as its human control should keep agent-written pull requests reviewable, for example by limiting them to a size that one reviewer can read and by requiring a second reviewer when a change involves one of the choices named above. The distinction between technical and decisional autonomy \cite{Gounaris2026} suggests a second measure: separate what an agent may do alone from what it may only propose.

\subsubsection{Instructions and checks} Developers already write ethical statements into agent configuration files \cite{gao2026ethicsconfigurationaiethics}, but whether agents follow them is unknown. A sentence in \texttt{AGENTS.md} therefore records what a team intends and does not enforce it. One paper in our corpus observes that strict norms against deception and manipulation ``that are unenforceable for humans might be enforceable in AI'' \cite{Tarsney2025}. Although Tarsney does not discuss SE workflows, one possible translation is to implement machine-checkable constraints as pre-merge checks; though many constraints cannot be checked in this way and these remain choices referred to a human.

\subsubsection{Changes to tests} A change that passes all tests is not necessarily acceptable, and the tests can be what the agent changed. Von Arx et al. report that frontier models modified tests or scoring code to raise their scores \cite{metr-2025-recent-reward-hacking}. Whether this is deception in the philosophical sense is the fifth question on our agenda, but the consequence for a team does not depend on the answer: passing tests provide weaker evidence when an agent can also modify the tests or verification mechanisms used to assess its work. Reviewers should check whether an agent-written change edits tests, build scripts, or other checks, and teams can require a named human owner for those files.

\subsubsection{Records of responsibility} Tweed \cite{Tweed2027} discusses how the layered ecosystem of agentic AI complicates the attribution of responsibility across the actors involved in developing and deploying these systems. A team cannot settle these questions for the whole ecosystem, but it can record, for each agent-written change, which agent and configuration produced the change and who approved it. Incident reports can then name that person and that configuration. A phrase such as ``the agent decided'' describes the agent in intentional terms, although ``agent'' in SE is a technical term that does not signal ``any philosophical notion of moral or epistemic agency'' \cite{Mussgnug2025}. Such a phrase names no accountable person.

\subsection{Threats to validity}
Following the framework of Wohlin et al.~\cite{WohlinValidity2024}, we now reflect on our study's validity. 

\subsubsection{Construct validity} The search string may miss relevant terms beyond \texttt{("ai agent" | "agentic ai")}, given that ``agentic AI'' is a recent and unstable label for this emerging trend.  

\subsubsection{Internal validity} We acknowledge that qualitative analysis is subject to researcher bias, which was mitigated by independent code checking and author discussion across different disciplines, resulting in multiple iterations. 

\subsubsection{External validity} For the current study, only PhilPapers was searched, and therefore ethics work published in computing, law, or science and technology studies (STS) venues may be missed; only English-language works were included in our analysis, which might omit philosophical traditions in other languages (such as Asian philosophy or Ubuntu philosophy). The implications in Section~\ref{ss:Implications} transfer arguments from other settings to SE. The quoted papers discuss negotiation chatbots \cite{Weirup2026}, AI companions \cite{Clowes2026}, and AI agents in research \cite{Hosseini2026}, not coding agents, and no practitioner has reviewed the implications. We therefore present them as options for teams and tie each one to an open question on our agenda. 

\subsubsection{Reliability} Thematic analysis involves interpretation, and the initial coding (iteration \#1), via the final \textit{kanban} visualiser, is released in the replication package. 

\section{Conclusion}\label{s:Conclusion}
This paper bridged the disciplinary gap by bringing the philosophical literature on the ethics of AI agents to SE. We conducted a systematic mapping study, reviewing $N=33$ peer-reviewed ethics papers indexed on PhilPapers (against a wider library of $140$ philosophical studies pertaining to AI agents/agentic AI), coded them thematically, and translated the resulting themes into a research agenda. We identified eight themes, of which the most prominent concerns echoed the principles on which AI ethics guidelines converge~\cite{Jobin2019}. Less frequent codes, such as ethical fading and deference to agents, are the most directly applicable to SE, and deserve future exploration. Our contribution to SE research is a thematic map of these concerns and { seven research questions} that translate them into SE terms, from  ethical instructions in agent configuration files to the effect of describing agent actions in intentional terms. Our contribution to SE practice is five implications for teams that use coding agents, which cover the choices an agent must refer to a human, the review of agent-written pull requests, checks that back up configuration-file instructions, changes to tests, and records of who approved what. Future work includes empirical validation of the agenda and the implications with SE practitioners, and extending the review beyond PhilPapers and English-language sources.

\section*{Data availability}\label{s:DataAvailability}
{
Replication package is available at \url{https://doi.org/10.5281/zenodo.24411462}.
}

\section*{Acknowledgments}
The authors acknowledge the use of Anthropic Claude (\textsc{Opus-5.5-Extra}) in this manuscript for (a) light copyediting, (b) generating figures as attributed in the captions, and (c) generating the boilerplate structure. In the research process, Claude was used to (a) markup/annotate longer PDFs for ease of comprehension, and (b) generate a thematic \textit{kanban} from \textsc{Atlas.TI} data to facilitate analysis of the results. 
Marc acknowledges the institutional support provided by Google at CAIDE.

\bibliography{bib}
\bibliographystyle{IEEEtran}
 
\end{document}